\documentstyle[sprocl]{article}

\input{psfig}

\begin{document}

{\noindent\small UNITU--THEP--9/1998 \hfill FAU--TP3--98/8 }
\vspace{5mm}

\title{A Solution to Coupled Dyson--Schwinger Equations for Gluons and Ghosts
       in Landau Gauge}

\author{A.~Hauck, R.~Alkofer}

\address{Universit\"at T\"ubingen, Institut f\"ur Theoretische Physik,
         Auf der Morgenstelle 14 \\
	 72076 T\"ubingen, Germany \\
	 E-mail: Andreas.Hauck@uni-tuebingen.de \\
         E-mail: Reinhard.Alkofer@uni-tuebingen.de} 

\author{L.~von Smekal}

\address{Universit\"at Erlangen--N\"urnberg, Institut f\"ur Theoretische
         Physik III, Staudtstr.~7 \\
         91058 Erlangen, Germany \\
	 E-mail: smekal@theorie3.physik.uni-erlangen.de}

\maketitle

\abstracts{ A truncation scheme for the Dyson--Schwinger equations of QCD in
Landau gauge is presented which implements the Slavnov--Taylor identities for
the 3--point vertex functions. Neglecting contributions from 4--point
correlations such as the 4--gluon vertex function and irreducible scattering
kernels, a closed system of equations for the propagators is obtained. For the
pure gauge theory without quarks this system of equations for the propagators
of gluons and ghosts is solved in an approximation which allows for an analytic
discussion of its solutions in the infrared: The gluon propagator is shown to
vanish for small spacelike momenta whereas the ghost propagator is found to be
infrared enhanced. The running coupling of the non--perturbative subtraction
scheme approaches an infrared stable fixed point at a critical value of the
coupling, $\alpha_c \simeq 9.5$. The results for the propagators obtained here
compare favorably with recent lattice calculations. }

\section{Introduction}

The main objective of this talk is to present results for the basic QCD
propagators and for the running coupling. Knowledge of these quantities may
yield further insight into the physical nature of confinement. One appealing
possibility to describe confinement is to ascribe the suppression of the
emission of colored states from color--singlet states to infrared divergencies
in the elementary correlators of QCD.

Thus, to study the infrared behavior of QCD amplitudes non--perturbative
methods are required, and, since divergences are anticipated, a formulation in
the continuum is desirable. Both of these are provided by studies of truncated
systems of Dyson--Schwinger equations (DSEs), the equations of motion of QCD
Green's functions. Typically, for their truncation, additional sources of
information like the Slavnov--Taylor identities, entailed by gauge invariance,
are used to express vertex functions in terms of the elementary 2--point
functions, i.e., the quark, ghost and gluon propagators. Those propagators can
then be obtained as selfconsistent solutions to non--linear integral equations
representing a closed set of truncated DSEs. Some systematic control over the
truncating assumptions can be obtained by successively including higher
$n$--point functions in selfconsistent calculations, and by assessing their
influence on lower $n$--point functions in this way. Until recently all
solutions to truncated DSEs of QCD in Landau gauge, even in absence of quarks,
relyed on neglecting ghost contributions
completely~\cite{Man79,Atk81,Bro89,Hau96}. While this particular problem is
avoided in ghost free gauges such as the axial gauge, in studies of the gluon
DSE in this gauge~\cite{Bak81}, the possible occurrence of an independent
second term in the tensor structure of the gluon propagator has so far been
disregarded~\cite{Bue95}. In fact, if the complete tensor structure of the
gluon propagator in axial gauge is taken into account, one arrives at equations
of no less complexitiy than the ghost--gluon system in the Landau gauge.

In this talk a simultaneous solution of a truncated set of DSEs for the
propagators of gluons and ghosts in Landau gauge is
presented~\cite{Sme97,Sme98}. An extension to this selfconsistent framework to
include quarks is possible and subject to further studies. The behavior of the
solutions in the infrared, implying the existence of a fixed point at a
critical coupling $\alpha_c \approx 9.5$, is obtained analytically. The gluon
propagator is shown to vanish for small spacelike momenta in the present
truncation scheme. This behavior, though in contradiction with previous DSE
studies~\cite{Man79,Atk81,Bro89,Hau96}, can be partially understood from the
observation that, in our present calculation, the previously neglected ghost
propagator assumes an infrared enhancement similar to what was then obtained
for the gluon.

In the meantime the qualitative behavior obtained in the studies to be reported
here is supported by investigations of the coupled gluon ghost DSEs using bare
vertices with~\cite{Atk97} and without~\cite{Atk98} angle approximation. As
expected however, the details of the result depend on the approximations
employed.

\section{The set of truncated gluon and ghost DSEs}

Besides all elementary 2--point functions, i.e., the quark, ghost and gluon
propagators, the DSE for the gluon propagator also involves the 3-- and
4--point vertex functions which obey their own DSEs. These equations involve
successively higher n--point functions. A first step towards a truncation of
the gluon equation is to neglect all terms with 4--gluon vertices. These are
the momentum independent tadpole term, an irrelevant constant which vanishes
perturbatively in Landau gauge, and explicit 2--loop contributions to the gluon
DSE. The latter are subdominant in the ultraviolet and will thus not affect the
behavior of the solutions for asymptotically high momenta. In the infrared it
has been argued that the singularity structure of the 2--loop terms does not
interfere with the one--loop terms~\cite{Vac95}. Without contributions from
4--gluon vertices (and quarks) the renormalized equation for the inverse gluon
propagator in Euclidean momentum space is given by
\begin{eqnarray}
  D^{-1}_{\mu\nu}(k)
  &=& Z_3 \, {D^{\hbox{\tiny tl}}}^{-1}_{\mu\nu}(k) \,
  + g^2 N_c\, Z_1  \frac{1}{2} \int {d^4q\over (2\pi)^4} \nonumber \\
  && \hskip -1cm \times \, \Gamma^{\hbox{\tiny tl}}_{\mu\rho\alpha}(k,-p,-q)
  \, D_{\alpha\beta}(q) D_{\rho\sigma}(p) \, \Gamma_{\beta\sigma\nu}(q,p,-k)
  \nonumber \\
  && \hskip -1.6cm - g^2 N_c \, \widetilde{Z}_1 \int \frac{d^4q}{(2\pi)^4} \;
  i q_\mu \, D_G(p)\, D_G(q)\, G_\nu(-q,p) \; ,
  \label{glDSE}
\end{eqnarray}
where we use positive definite metric, $g_{\mu\nu} = \delta_{\mu\nu}$. Color
indices are suppressed and the number of colors is fixed, $N_c = 3$. Furthermore
$p = k - q$, $D^{\hbox{\tiny tl}}$ and $\Gamma^{\hbox{\tiny tl}}$ are
the tree level propagator and 3--gluon vertex, $D_G$ is the ghost
propagator and $\Gamma$ and $G$ are the fully dressed 3--point
vertex functions. The equation for the ghost propagator in Landau gauge QCD,
without any truncations, is given by
\begin{eqnarray}  
  D_G^{-1}(k) &=&
    -\widetilde{Z}_3 \, k^2 \, +\, g^2 N_c \, \widetilde{Z}_1 \label{ghDSE} \\
  && \hskip -1cm \times \int {d^4q\over (2\pi)^4} \;
  i k_\mu \, D_{\mu\nu}(k-q) \, G_\nu(k,q) \, D_G(q) \; . \nonumber
\end{eqnarray}
The renormalized propagators for ghosts and gluons and the renormalized
coupling are defined from the respective bare quantities by  introducing
multiplicative renormalization constants, $\widetilde Z_3 D_G := D^0_G$, $Z_3
D_{\mu\nu} :=  D^0_{\mu\nu}$ and $Z_g g := g_0$. Furthermore, $Z_1 = Z_g
Z_3^{3/2}$, $\widetilde Z_1 = Z_g Z_3^{1/2} \widetilde Z_3$, and we use that
$\widetilde Z_1 = 1$ in Landau gauge. The ghost and gluon propagators are
parameterized by their respective renormalization functions $G$ and $Z$,
\begin{equation}
  D_G(k) = -\frac{G(k^2)}{k^2} \; , \quad
  D_{\mu\nu}(k) = \bigg( \delta_{\mu\nu} - \frac{k_\mu k_\nu}{k^2}\bigg)
                  \frac{Z(k^2)}{k^2} \; . 
  \label{rfD}
\end{equation}
In order to arrive at a closed set of equations for the functions $G$ and $Z$,
we use a form for the ghost--gluon vertex which is based on a construction from
its Slavnov--Taylor identity (STI) which can be derived from the usual
Becchi--Rouet--Stora invariance neglecting irreducible 4--ghost correlations in
agreement with the present level of truncation~\cite{Sme98}. This together with
the symmetry of the ghost--gluon vertex fully determines its form at the
present level of truncation. There are no undetermined transverse terms in this
case:
\begin{equation}
  G_\mu(p,q) =
    i q_\mu \, \frac{G(k^2)}{G(q^2)}
    + i p_\mu \, \biggl( \frac{G(k^2)}{G(p^2)} - 1 \biggr) .
  \label{fvs}
\end{equation}
With this result, we can construct the 3--gluon vertex according to procedures
developed and used previously~\cite{Bar80},
\begin{eqnarray} 
  \Gamma_{\mu\nu\rho}(p,q,k) &=&  
  \frac{1}{2} A_+(p^2,q^2;k^2) \, \delta_{\mu\nu} \, i (p-q)_\rho\,
  \label{3gv} \\
  && \hskip -2cm + \frac{1}{2}  A_-(p^2,q^2;k^2) \, \delta_{\mu\nu} i (p+q)_\rho
  + \frac{A_-(p^2,q^2;k^2)}{p^2-q^2} 
  \nonumber \\
  && \hskip -2cm \times ( \delta_{\mu\nu} pq - p_\nu q_\mu) \,
  i (p-q)_\rho + \hbox{cyclic permutations} \; , \nonumber 
\end{eqnarray} 
\begin{displaymath}
   A_\pm (p^2,q^2;k^2) = \frac{G(k^2) G(q^2)}{G(p^2)Z(p^2)}
    \pm \frac{ G(k^2) G(p^2)}{G(q^2)Z(q^2)} \; .
\end{displaymath}
Some additionally possible terms, transverse with respect to all three gluon
momenta, cannot be constrained by its STI and are thus disregarded.


We solve the coupled system of integral equations of the present truncation
scheme in an one--dimen\-sion\-al approximation:
For integration momenta $q^2 > k^2$ we use the perturbative anticipation
that the functions $Z$ and $G$ are slowly varying with their 
arguments, and that we may thus replace all arguments by the integration
momentum $q^2$. This assumption ensures the correct leading perturbative
behavior at one--loop level~\cite{Sme98}. An analogous assumption underlies
the Mandelstam approximation~\cite{Man79,Atk81,Bro89,Hau96}. 
Infrared enhanced solutions tend to invalidate this assumption, in
particular for small $q^2 < k^2$, however. For integration momenta $q^2 < k^2
$ we therefore use the angle approximation replacing $G((k-q)^2) \to  G(k^2)$
and similarly for $Z$, which preserves the limit of the integrands
for $q^2 \to 0$. The DSEs (\ref{glDSE}) and (\ref{ghDSE}) then simplify to 
\begin{eqnarray}
  \frac{1}{Z(k^2)} &=&
    Z_3 + Z_1 \frac{g^2}{16\pi^2} \Bigg\{ \int_{0}^{k^2} \frac{dq^2}{k^2} \,
      \left( \frac{7}{2}\frac{q^4}{k^4} - \frac{17}{2}\frac{q^2}{k^2}
             - \frac{9}{8} \right) Z(q^2) G(q^2)  \nonumber \\
    && \hspace{20mm}
      + \int_{k^2}^{\Lambda_{\hbox{\tiny{UV}}}^2} \frac{dq^2}{q^2} \,
    \left( \frac{7}{8} \frac{k^2}{q^2} - 7\right) Z(q^2)
    G(q^2) \Bigg\}  \nonumber \\
    && \hspace{-10mm} + \frac{g^2}{16\pi^2} \Bigg\{ \int_{0}^{k^2}
    \frac{dq^2}{k^2} \frac{3}{2} \frac{q^2}{k^2} G(k^2) G(q^2) -
    \frac{1}{3} G^2(k^2) + \frac{1}{2} \int_{k^2}^{\Lambda_{\hbox{\tiny{UV}}}^2}
    \frac{dq^2}{q^2} \, G^2(q^2) \Bigg\} \; , \label{odZDSE}  \\
 \frac{1}{G(k^2) } &=&
   \widetilde{Z}_3 - \frac{g^2}{16\pi^2} \, \frac{9}{4}
   \Bigg\{ \, \frac{1}{2} \, Z(k^2) G(k^2)
   + \int_{k^2}^{\Lambda_{\hbox{\tiny{UV}}}^2} \frac{dq^2}{q^2} \,
   Z(q^2) G(q^2) \Bigg\}  \; .
   \label{odGDSE}
\end{eqnarray}
We introduced an $O(4)$--invariant momentum cutoff $\Lambda_{\hbox{\tiny{UV}}}$
to account for logarithmic ultraviolet divergences which are absorbed by the
renormalization constants $Z_3$ and $\widetilde{Z}_3$. The details of the
renormalization and the numerical procedure are given elsewhere~\cite{Sme98}.

The angle approximation as we use it for small integration momenta was also
used in a very recent study of the coupled system of ghost and gluon DSEs with
bare vertices~\cite{Atk97}. However, using this approximation for arbitrary
$q^2$ (i.e.\ also for $q^2 > k^2$) one does not recover the renormalization
group improved one--loop results for asymptotically large
momenta~\cite{Atk97}. 

For a reduced set of equations the effect of the angle approximation has been
demonstrated to be only quantitative, the qualitative infrared behavior of the
solution remaining the same~\cite{Atk98}. It will nevertheless be important to
assess the sensitivity of the results to the modified angle approximation
further in future. 

To deduce the infrared behavior of the propagators here, we make the Ansatz
that for $x := k^2 \to 0$ the product $Z(x)G(x) \to c x^\kappa$ with $\kappa
\not= 0$ and some constant $c$. The special case $\kappa = 0$ leads to a
logarithmic singularity in Eq.~(\ref{odGDSE}) for $x \to 0$ which precludes
the possibility of a selfconsistent solution. In order to obtain a positive
definite function $G(x)$ for positive $x$ from an equally positive $Z(x)$, as
$x\to 0$, we obtain the further restriction $0 < \kappa <
2$. Eq.~(\ref{odGDSE}) then yields, 
\begin{eqnarray}  
  G(x) &\to &  \left( g^2\gamma_0^G \left(\frac{1}{\kappa} - \frac{1}{2}
  \right) \right)^{-1}  c^{-1} x^{-\kappa} \quad \Rightarrow \label{loirG} \\
  Z(x) & \to &  \left( g^2\gamma_0^G \left(\frac{1}{\kappa} - \frac{1}{2}
  \right) \right) \, c^{2} x^{2\kappa}   \; ,
  \label{loirZ1}
\end{eqnarray}
where $\gamma_0^G = 9/(64\pi^2)$ is the leading order perturbative coefficient
of the anomalous dimension of the ghost field. Using (\ref{loirG}) and
(\ref{loirZ1}) in Eq.~(\ref{odZDSE}), we find that the 3--gluon loop
contributes terms $\sim  x^\kappa$ to gluon equation for $x \to 0$ while the
dominant (infrared singular) contribution arises from the ghost--loop, i.e.,
\begin{displaymath}
  Z(x) \to g^2\gamma_0^G \, \frac{9}{4} \left(\frac{1}{\kappa} -
  \frac{1}{2} \right)^2 \! \left( \frac{3}{2}\, \frac{1}{2-\kappa} -
  \frac{1}{3} + \frac{1}{4\kappa} \right)^{-1}\!\! c^2 x^{2\kappa} . 
\end{displaymath}
Comparing this to (\ref{loirZ1}) we obtain a quadratic equation for $\kappa$
with a unique solution for the exponent in $0 < \kappa < 2$:
\begin{equation}
  \kappa = \frac{61-\sqrt{1897}}{19} \simeq 0.92 \; .
\end{equation}
The leading behavior of the gluon and ghost renormalization
functions and thus of their propagators is entirely due to ghost contributions.
The details of the approximations to the 3--gluon loop have no influence on the
above considerations. Compared to the Mandelstam approximation, in which the 
3--gluon loop alone determines the infrared behavior of the gluon propagator
and the running coupling in Landau gauge~\cite{Man79,Atk81,Bro89,Hau96}, this
shows the importance of ghosts. The result presented here implies an infrared
stable fixed point in the non--perturbative running coupling of our subtraction
scheme, defined by 
\begin{equation}
  \alpha_S(s) = \frac{g^2}{4\pi} Z(s) G^2(s)
    \quad \to \quad
    \frac{16\pi}{9} \left(\frac{1}{\kappa} - \frac{1}{2}\right)^{-1}
    \approx 9.5 \; , 
\end{equation} 
for $s\to 0$. This is qualitatively different from the infrared singular 
coupling of the Mandelstam approximation~\cite{Hau96}.

\begin{figure}[t]
  \centerline{ \psfig{figure=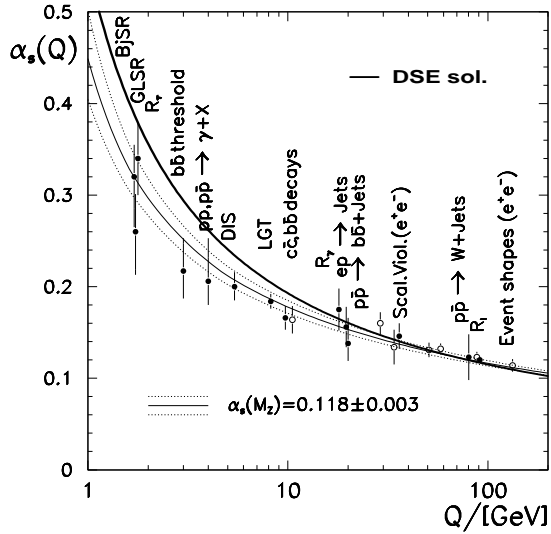,width=.6\textwidth} }
  \caption{ The present result compared to the {\it
  world's} experimental data according to ref.~\protect\cite{Sch97}. }
  \label{Fig:alpha_s}
\end{figure}

The momentum scale in our calculations is fixed from the phenomenological value
$\alpha_S (M_Z) = 0.118$ at the mass of the $Z$--boson. The ratio of the $Z$--
to the $\tau$--mass, $M_Z/M_\tau \simeq 51.5$, then yields $\alpha_S(M_\tau) =
0.38$. In Fig.~\ref{Fig:alpha_s} we compare our result for the running
coupling to the experimental data summarized in fig. 6 of
ref.~\cite{Sch97}. The enhancement at lower momenta can be attributed to the
omission of quarks in our calculation ($N_f=0$).

\section{Comparison to lattice results}

\begin{figure}[t]
  \centerline{ \psfig{figure=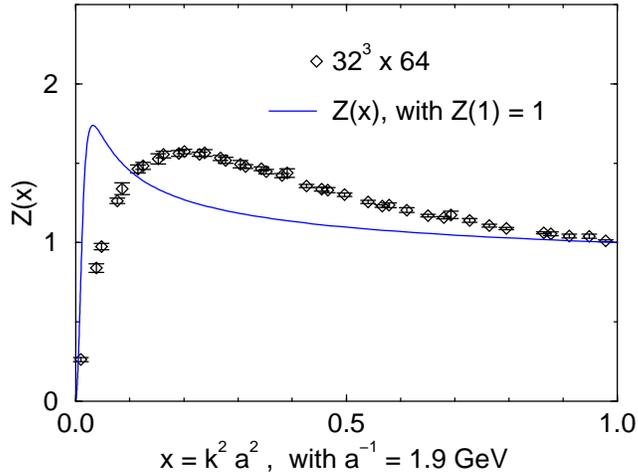,width=.7\textwidth} }
  \caption{ The numerical result for the gluon propagator from Dyson--Schwinger
            equations (solid line) compared to recent lattice
            calculations~\protect\cite{Der98}. }
  \label{Fig:Gluon-Lattice}
\end{figure}

It is interesting to compare our solutions to recent lattice results available
for the gluon propagator~\cite{Der98} and for the ghost propagator~\cite{Sum96}
using lattice versions to implement the Landau gauge condition. In
Fig.~\ref{Fig:Gluon-Lattice} we compare our solution for the gluon propagator
to the most recent data for the lattice gluon correlator~\cite{Der98}.

\begin{figure}[t]
  \centerline{ \psfig{figure=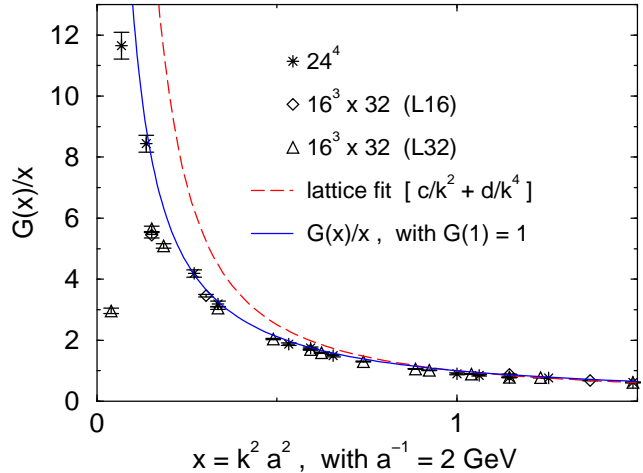,width=.7\textwidth} }
  \caption{ The numerical result for the ghost propagator from Dyson--Schwinger
            equations (solid line) compared to lattice data from Fig.~1 in
            in~\protect\cite{Sum96} and a fit as given by these authors. }
  \label{Fig:Ghost-Lattice}
\end{figure}

In Fig.~\ref{Fig:Ghost-Lattice} we compared our infrared enhanced ghost
propagator to corresponding recent results~\cite{Sum96}. It is quite amazing to
observe that our solution fits the lattice data at low momenta significantly
better than the fit to an infrared singular form $D_G(k^2) = c/k^2 + d/k^4$
given in~\cite{Sum96}. We therefore conclude that present lattice calculations
confirm the existence of an infrared enhanced ghost propagator of the form $D_G
\sim 1/(k^2)^{1+\kappa}$ with $0 < \kappa < 1$. This is an interesting result
for yet another reason: In this calculation the Landau gauge condition was
supplemented by an algorithm to select  gauge field configurations from the
fundamental modular region which is to avoid Gribov copies. Thus, our results
suggest that the existence of such copies of gauge configurations might have
little effect on the solutions to Landau gauge DSEs. This could also explain
the similarity of our solutions to the infrared behavior obtained by
D.~Zwanziger~\cite{Zwa94}.

\section{Summary}

In summary, we presented a solution to a truncated set of coupled
Dyson--Schwinger equations for gluons and ghosts in Landau gauge. The infrared
behavior of this solution, obtained analytically, represents a strongly
infrared enhanced ghost propagator and an infrared vanishing gluon propagator.

Using bare vertices qualitatively similar results are obtained in an
investigation of the coupled gluon and ghost DSEs. Relying on an angle
approximation one obtains $\kappa \simeq 0.77$ for the exponent of the infrared
leading term~\cite{Atk97} whereas without angle approximation the exponent
changes to $\kappa = 1$~\cite{Atk98}.

A highly infrared singular behavior for the ghost propagator in Landau gauge
has also been suggested from studies of the influence of a complete gauge
fixing~\cite{Zwa94}. Our results for the propagators, in particular for the
ghost, compare favorably with recent lattice calculations~\cite{Der98,Sum96}.

The Euclidean gluon correlation function presented here can be shown to violate
reflection positivity~\cite{Sme98}, which is a necessary and sufficient
condition for the existence of a Lehmann representation~\cite{Fritz}. We
interpret this as representing confined gluons. In order to understand how
these correlations can give rise to confinement of quarks, it will be necessary
to include the quark propagator.

The size of the coupling at the fixed point, $\alpha_c \approx 9.5$, is
however, a good indication that dynamical chiral symmetry breaking will be
generated in the quark DSE. The existence such an infrared fixed point is in
qualitative disagreement with previous studies of the gluon DSE neglecting
ghost contributions in Landau gauge~\cite{Man79,Atk81,Bro89,Hau96}. This shows
that ghosts are important, in particular, at low energy scales relevant to
hadronic observables.

\section*{Acknowledgments}

A.~Hauck wants to thank the organizers of the workshop and gratefully
acknowledges the great hospitality experienced in Adelaide.

We thank D.~Leinweber, J.~I.~Skullerud and A.~G.~Williams for communicating
their results for the lattice gluon propagator to us. L.~v.~Smekal thanks the
Argonne National Laboratory, where large part of this work was accomplished.
This work was supported by DFG under contract Al 279/3-1, by the
Gra\-duier\-ten\-kolleg T\"ubingen, BMBF Erlangen co.~\#~06--ER--809 and the
US-DOE, Nuclear Physics Division, contract W-31-109-ENG-38.


\section*{References}

\begin{thebibliography}{99}
\bibitem{Man79}
  S.~Mandelstam, Phys.~Rev.~D {\bf 20}, 3223 (1979).
\bibitem{Atk81}
  D.~Atkinson {\it et al.}, J.~Math.~Phys.~{\bf 22}, 2704 (1981);
  D.~Atkinson, P.~W.~Johnson and K.~Stam, {\it ibid.} {\bf 23}, 1917 (1982).
\bibitem{Bro89}
  N.~Brown and M.~R.~Pennington, Phys.~Rev.~D {\bf 39}, 2723 (1989).
\bibitem{Hau96}
  A.~Hauck, L.~v.~Smekal and R.~Alkofer, {\it e-print} hep--ph/9604430
\bibitem{Bak81}
  M.~Baker, J.~S.~Ball and F.~Zachariasen, Nucl.\ Phys.\ {\bf B186}, 531/560
  (1981); W.~J.~Schoenmaker, {\it ibid}, {\bf B194}, 535 (1982); J.~R.~Cudell
  and D.~A.~Ross, {\it ibid}, {\bf B358}, 247 (1991).
\bibitem{Bue95}
  K.~B\"uttner and M.~R.~Pennington, Phys.~Rev.~D {\bf 52}, 5220 (1995).
\bibitem{Sme97}
  L.~v.~Smekal, A.~Hauck and R.~Alkofer, Phys.~Rev.~Lett.\ {\bf 79},
  3591 (1997).
\bibitem{Sme98}
  L.~v.~Smekal, A.~Hauck and R.~Alkofer, Ann.~Phys.\ to be published;
  {\it e-print} hep--ph/9707327.
\bibitem{Atk97}
  D.~Atkinson and J.~Bloch, {\it e-print} hep--ph/9712459 and contributions
  to these proceedings.
\bibitem{Atk98}
  D.~Atkinson and J.~Bloch, {\it e-print} hep--ph/9802239 and contributions
  to these proceedings.
\bibitem{Vac95}
  L.~G.~Vachnadze, N.~A.~Kiknadze and A.~A.~Khelashvili,
  Theor.\ Math.\ Phys.~{\bf 102}, 34 (1995).
\bibitem{Bar80}
  U.~Bar--Gadda, Nucl. Phys. {\bf B163},
  312 (1980; S.~K.~Kim and M.~Baker, {\it ibid.} {\bf B164}, 152 (1980);
  J.~S.~Ball and {T.-W.~Chiu}, Phys.~Rev.~D {\bf 22}, 2550 (1980).
\bibitem{Sch97}
  M.~Schmelling, {\it e-print} hep--ex/9701002.
\bibitem{Der98}
  D.~Leinweber, J.~I.~Skullerud and A.~G.~Williams, private communications;
  hep--lat/9803015 and contribution to these proceedings.
\bibitem{Sum96}
  H.~Suman and K.~Schilling, Phys.~Lett.~{\bf B373}, 314 (1996).
\bibitem{Zwa94}
  D.~Zwanziger, Nucl.~Phys.\ {\bf B378}, 525 (1992);
  {\it ibid.} {\bf B412}, 657 (1994)].
\bibitem{Fritz}
  F.~Coester, private communication.
\end{thebibliography}
\end{document}